\documentclass[12pt]{article}
\usepackage[mathscr]{eucal}
\usepackage{amssymb,latexsym}
\usepackage{verbatim}
\newtheorem{thm}{Theorem}[section]
\newcommand{\qed}{\hfill $\Box$ \medskip}

\newcommand\calT{{\mathcal{T}}}

\renewcommand\l{\lambda}

\renewcommand\S{\Sigma}

\renewcommand\d{\partial}

\renewcommand\L{\triangle}
\newcommand\D{\nabla}
\newcommand\e{\epsilon}

\newcommand\la{\langle}
\newcommand\ra{\rangle}

\renewcommand\l{\lambda}

\renewcommand\th{\theta}

\newcommand\<{\la}
\renewcommand\>{\ra}

\newcommand\beq{\begin{eqnarray}}
\newcommand\eeq{\end{eqnarray}}
\newcommand\ben{\begin{enumerate}}
\newcommand\een{\end{enumerate}}
\newcommand\bit{\begin{itemize}}
\newcommand\eit{\end{itemize}}

\newcounter{mnotecount}[section]

\title{A generalization of Hawking's black hole topology\\ theorem to higher dimensions}

\author{Gregory J. Galloway
 \\Department of Mathematics \\ University of Miami
 \\ \\ Richard Schoen \\ Department of Mathematics\\Stanford University \\}

\begin{document}
\date{}
\maketitle

\begin{abstract}
Hawking's theorem on the topology of black holes asserts that
cross sections of the event horizon in 4-dimensional
asymptotically flat stationary black hole spacetimes obeying the
dominant energy condition are topologically 2-spheres.  This
conclusion extends to outer apparent horizons in spacetimes that
are not necessarily stationary.  In this paper we obtain a natural
generalization of Hawking's results  to higher dimensions by
showing that cross sections of the event horizon (in the
stationary case) and outer apparent horizons (in the general case)
are of positive Yamabe type, i.e., admit metrics of positive
scalar curvature.  This implies many well-known restrictions on
the topology, and is consistent with recent examples of five
dimensional stationary black hole spacetimes with horizon topology
$S^2 \times S^1$.  The proof is inspired by previous work of
Schoen and Yau on the existence of solutions to the Jang
equation (but does not make direct use of that equation).
\end{abstract}

\newpage

\section{Introduction}

A basic result in the theory of black holes is Hawking's
theorem \cite{H1, HE} 
on the topology of black holes, which asserts that cross sections
of the event horizon in  $4$-dimensional asymptotically flat stationary black hole spacetimes obeying the dominant energy condition are spherical (i.e., topologically
$S^2$).  The proof is a beautiful variational argument, showing that
if a cross section  has genus $\ge 1$ then it can be deformed along
a null hypersurface to an
outer trapped surface outside of the event horizon, which is forbidden
by standard results on black holes \cite{HE}.\footnote{Actually the torus $T^2$ arises as a borderline case 
in Hawking's argument, but can occur only under special circumstances.} 
In \cite{H2}, Hawking showed that his black hole topology result extends,
by a similar argument,
to outer apparent horizons in black hole spacetimes that are not necessarily stationary.
(A related result had been shown by Gibbons \cite{Gi} in the time-symmetric case.)
Since Hawking's arguments rely on the Gauss-Bonnet theorem, these results
do not directly extend to higher dimensions.  

Given the current interest in higher dimensional black holes, it is of interest
to determine which properties of black holes in four spacetime dimensions
extend to higher
dimensions.  In this note we obtain a natural generalization of Hawking's
theorem on the topology of black holes to higher dimensions.  The conclusion in
higher dimensions is not that the horizon topology is spherical; that would
be too strong, as evidenced by the striking example of Emparan and Reall
\cite{ER} of a  stationary vacuum black hole spacetime in five dimensions
with horizon topology $S^2 \times S^1$.   The natural conclusion in higher
dimensions is that cross sections of the event horizon (in the stationary case),
and outer apparent horizons (in the general case) are of positive
Yamabe type, i.e. admit
 metrics of positive scalar curvature.  
As noted in \cite{CG}, in the time symmetric case this follows from the minimal surface methodology  of Schoen and Yau \cite{SY2} in their treatment of manifolds of positive
scalar curvature.  The main point of the present paper is to show that this conclusion
 remains valid without any condition on the extrinsic curvature of space.  
 That such a result might be expected to hold
 is suggested by work in \cite[Section 4]{SY3}, which implies
 that the apparent horizons corresponding to  the blow-up of solutions
 of the Jang equation, as described
 in \cite{SY3}, are of positive Yamabe type.  We emphasize, however, that we do not need to make use of the Jang equation here.\footnote{
 In any
case, the parametric estimates of \cite{SY3} which are used to construct
solutions of the Jang equation asymptotic to vertical cylinders over
apparent horizons are generally true only in low dimensions.} 
 
 Much is now known about  the topological obstructions to the existence of 
 metrics of positive scalar curvature in higher dimensions.  
 While the first major result along these
 lines is the famous theorem of Lichnerowicz \cite{Li} concerning the vanishing
 of the $\hat A$ genus, a key advance in our understanding 
 was made in the late 70's and early 80's by 
 Schoen and Yau \cite{SY1,SY2}, and Gromov and Lawson \cite{GL1,GL2}.
 A brief review of these results, relevant to the topology
 of black holes,  was considered in \cite{CG}.  We shall recall the situation
 in five spacetime dimensions in the next section, after the 
 statement of our main result.

\section{The main result}

Let $V^n$ be an $n$-dimensional, $n \ge 3$, spacelike
hypersurface in a spacetime $(M^{n+1},g)$.   Let $\S^{n-1}$ be a closed
hypersurface in $V^n$, and assume that $\S^{n-1}$ separates
$V^n$ into an ``inside" and an ``outside".   Let $N$ be the outward unit normal
to $\S^{n-1}$ in $V^n$, and let $U$ be the future directed  unit normal
to $V^n$ in $M^{n+1}$.  Then $K = U +N$ is an outward null normal  field
to $\S^{n-1}$, unique up to scaling.  

The null second fundamental form of $\S$ with respect to $K$ is, for each $p \in \S$,
the bilinear form defined by,
\beq
\chi : T_p\S \times T_p\S \to \Bbb R , \qquad \chi(X,Y) = \<\D_XK, Y\>  \,,
\eeq
where $\<\,,\,\>=g$  and $\D$ is the Levi-Civita connection,
 of $M^{n+1}$.  Then the null expansion of $\S$ is defined as 
$\theta = {\rm tr}\, \chi = h^{AB}\chi_{AB} = {\rm div}\,_{\S} K$, where
$h$ is the induced metric on $\S$.

We shall say $\S^{n-1}$ is an {\it outer apparent horizon} in $V^n$
provided, (i) $\S$ is marginally outer trapped, i.e., $\theta = 0$,
and (ii) there are no outer trapped surfaces  outside of $\S$.
The latter means  that there is no $(n-1)$-surface
$\S'$ contained in the region of $V^n$ outside of $\S$ which is homologous to $\S$
and which  has negative expansion
$\th < 0$ with respect to its outer null normal (relative to $\S$).
Heuristically, $\S$ is the ``outer limit" of outer trapped surfaces in $V$.

Finally, a spacetime $(M^{n+1},g)$ satisfying the Einstein equations
\beq
R_{ab} - \frac12 R g_{ab} = T_{ab} 
\eeq
is said to obey the dominant energy condition
provided the energy-momentum tensor $\calT$ satisfies
$\calT(X,Y) = T_{ab}X^aY^b \ge 0$ for all future pointing
causal vectors $X,Y$.

We are now ready to state the main theorem.

\begin{thm}\label{main}
Let $(M^{n+1},g)$, $n \ge 3$, be a spacetime satisfying the
dominant energy condition.  If $\S^{n-1}$ is an outer apparent horizon 
in $V^n$ then $\S^{n-1}$ is of positive Yamabe type, unless
$\S^{n-1}$ is Ricci flat (flat if $n =3,4$) in the induced metric, and 
both $\chi$ and  $\calT(U,K) = T_{ab}U^aK^b$ vanish on $\S$.
\end{thm}

Thus, except under special circumstances, $\S^{n-1}$ is of positive
Yamabe type.  As noted in the introduction, this implies various restrictions
on the topology of $\S$.  Let us focus on the case dim $M = 5$, and hence
dim $\S = 3$, and assume, by taking a double cover if necessary, that
$\S$ is orientable.  Then by well-known results of Schoen-Yau \cite{SY2}
and Gromov-Lawson \cite{GL2}, topologically, $\S$ must be a finite connected
sum of  spherical spaces (homotopy $3$-spheres, perhaps with 
identifications) and  $S^2 \times S^1$'s.  Indeed, by the prime decomposition
theorem, $\S$ can be expressed as a connected sum of spherical spaces,
$S^2 \times S^1$'s, and $K(\pi,1)$ manifolds (manifolds whose
universal covers are contractible).  But as $\S$ admits a metric of positive scalar
curvature, it cannot have any $K(\pi,1)$'s in its prime decomposition.  
Thus, the basic horizon topologies in dim $M=5$ are $S^3$ and 
$S^2 \times S^1$, both of which are realized by nontrivial black hole spacetimes.
Under stringent geometric assumptions on the horizon, a related conclusion
is arrived at in \cite{HOY}.

\medskip
\noindent
{\it Proof of the theorem:}  We consider normal variations of $\S$ in $V$, i.e., variations 
$t \to \S_t$ of $\S = \S_0$, $-\e <t < \e$, with variation vector field 
$V = \left . \frac{\d}{\d t}\right |_{t=0} = \phi N$,  $\phi \in C^{\infty}(\S)$.   
Let $\th(t)$ denote
the null expansion of $\S_t$ with respect to $K_t = U + N_t$, where $N_t$ is the
outer unit normal field to $\S_t$ in $V$.  A computation shows \cite{CG, AMS},
\beq\label{der}
\left . \frac{\d\th}{\d t} \right |_{t=0} = -\triangle \phi + 2\<X,\D\phi\>  + \left(Q+{\rm div}\, X - |X|^2 \right)\phi \,,
\eeq
where,
\beq\label{Q}
Q = \frac12 S - \calT(U,K) - \frac12 |\chi|^2   \,,
\eeq
$S$ is the scalar curvature of $\S$, $X$ is the vector field on $\S$
defined by $X = {\rm tan}\,(\D_NU)$, and $\<\,,\,\>$ now denotes
the induced metric $h$ on $\S$.

Introducing as in \cite{AMS} the operator 
$L = -\triangle + \<X, \D(\,)\> +(Q + {\rm div}\,X - |X|^2)$, 
Equation~(\ref{der}) may be expressed as,
\beq\label{der2}
\left . \frac{\d\th}{\d t} \right |_{t=0} = L(\phi)  \,.
\eeq
$L$ is the stability operator associated with variations in the null expansion $\th$.
In the time symmetric case the vector field $X$ vanishes, and $L$ reduces
to the classical stability operator of minimal surface theory, as expected \cite{CG}.

As discussed in \cite{AMS}, although $L$ is not in general self adjoint, its principal
eigenvalue $\lambda_1$ is real, and one can choose a principal eigenfunction
$\phi$ which is strictly positive, $\phi > 0$.  Using the eigenfunction $\phi$ to
define our variation, we have from (\ref{der2}),
\beq\label{der3}
\left . \frac{\d\th}{\d t} \right |_{t=0} =\l_1 \phi \,.
\eeq
The eigenvalue $\l_1$ cannot be negative, for otherwise (\ref{der3}) would
imply that  $\frac{\d\th}{\d t}< 0$ on $\S$.  Since $\th = 0$ on $\S$, this would mean
that for $t>0$ sufficiently small, $\S_t$ would be outer trapped, contrary to our assumptions.

Hence, $\l_1 \ge 0$, and we conclude for the variation determined by the
positive eigenfunction $\phi$ that $\left . \frac{\d\th}{\d t} \right |_{t=0} \ge 0$.
By completing the square on the right hand side of Equation (\ref{der}), this implies
that the following  inequality holds,
\beq
 -\triangle \phi +\left(Q+{\rm div}\, X  \right)\phi   +
\phi|\D \ln\phi|^2  - \phi|X - \D\ln\phi|^2  \ge 0.
\eeq
Setting $u = \ln \phi$, we obtain,
\beq\label{u-ineq}
-\triangle u +Q + {\rm div}\,X  - |X - \D u|^2 \ge 0 \,.
\eeq

As a side remark,  note that substituting the expression for 
$Q$ into (\ref{u-ineq}) and integrating gives that the total scalar curvature
of $\S$ is nonnegative, and in fact is positive, except under special circumstances. 
In four spacetime dimensions one may then apply the Gauss-Bonnet theorem to recover Hawking's
theorem; in fact this is essentially Hawking's original argument.  However, in higher
dimensions the positivity of the total scalar curvature, in and of itself,  does not provide any topological information.  

To proceed with the higher dimensional case, we first absorb the Laplacian term
$\triangle u = {\rm div}\,(\D u)$ in (\ref{u-ineq})  into the divergence term to
obtain,
 \beq
Q + {\rm div}\,(X- \D u)  - |X - \D u|^2 \ge 0  \, .
\eeq
Setting $Y = X - \D u$, we arrive at the inequality,
\beq
- Q + |Y|^2 \le {\rm div}\, Y  \,.
\eeq
Given any $\psi \in C^{\infty}(\S)$, we multiply  through by $\psi^2$ and derive,
\beq
-\psi^2 Q +\psi^2 |Y|^2 &\le& \psi^2 {\rm div}\, Y \nonumber\\
& = & {\rm div}\,(\psi^2Y) - 2\psi \< \D\psi,Y \>  \nonumber  \\
& \le &  {\rm div}\,(\psi^2Y) + 2|\psi| |\D \psi| |Y| \nonumber\\
& \le & {\rm div}\,(\psi^2Y) + |\D \psi|^2 + \psi^2|Y|^2  \, .
\eeq 

Integrating the above inequality yields,
\beq\label{psi-ineq}
\int_{\S} |\D \psi|^2 + Q \psi^2 \ge 0 \quad \mbox{for all } \psi \in C^{\infty}(\S)  \,,
\eeq 
where $Q$ is given in (\ref{Q}).

At this point rather standard arguments become applicable \cite{SY3, CG}.
Consider the eigenvalue problem,
\beq\label{eigen}
- \triangle\psi + Q\psi = \mu \psi  \,.
\eeq
Inequality (\ref{psi-ineq}) implies that the first eigenvalue $\mu_1$
of (\ref{eigen}) is nonnegative, $\mu_1 \ge 0$.  Let $f \in C^{\infty}(\S)$
be an associated eigenfunction; $f$ can be chosen to be strictly
positive.

Now consider $\S$ in the conformally related metric $\tilde h = f^{2/n-2} h$.
The scalar curvature $\tilde S$ of $\S$ in the metric
$\tilde h$ is given by,
\beq\label{scalar}
\tilde S & = & f^{-n/(n-2)}\left (-2\L f + S f + \frac{n-1}{n-2} \frac{|\D f|^2}{f}\right)
\nonumber  \\
& = &  f^{-2/(n-2)}\left (2\mu_1  + 2\calT(U,K) + |\chi|^2 + \frac{n-1}{n-2} 
\frac{|\D f|^2}{f^2} \right)\,,
\eeq
where, for the second equation, we have used (\ref{eigen}), with $\psi = f$, and 
(\ref{Q}).

Since, by the dominant energy condition, $\calT(U,K) \ge 0$, Equation (\ref{scalar})
implies that $\tilde S \ge 0$.  If $\tilde S > 0$ at some point, then by well known results
\cite{KW} one can conformally change $\tilde h$  to a metric 
of strictly positive scalar curvature, and the theorem follows.  If $\tilde S$ vanishes
identically then, by Equation (\ref{scalar}), $\mu_1 = 0$, $\calT(U,K) \equiv 0$, $\chi \equiv 0$ and $f$ is constant.  Equation (\ref{eigen}), with $\psi = f$ and Equation
(\ref{Q}) then  imply that $S \equiv  0$.  By a result of Bourguinon (see \cite{KW}),
it follows that $\S$ carries a metric of positive scalar curvature unless  it is
Ricci flat.  The theorem now follows. \qed

\medskip
\noindent
{\bf Concluding Remarks.}

\smallskip
\noindent
1. Let $\S^{n-1}$ be a closed $2$-sided hypersurface in the spacelike
hypersurface $V^n \subset M^{n+1}$.  Then there exists a neighborhood
$W$ of $\S^{n-1}$ in $V^n$ such that $\S$ separates $W$ into an
``inside" and an ``outside".  Suppose $\S$ is marginally outer  trapped,
i.e., $\theta = 0$ with respect to the outer null normal to $\S$.  
Following the
terminology introduced in \cite{AMS}, we say that $\S$ is stably outermost
(respectively, strictly stably outermost) provided the principal eigenvalue  $\l_1$ of 
the stability operator $L$  introduced in \ref{der2} satisfies $\l_1 \ge 0$ 
(resp., $\l_1 >0$).  It is clear from the proof  that
the conclusion of Theorem \ref{main} 
remains valid for marginally outer trapped surfaces  $\S$ that are stably outermost.   Moreover the conclusion that $\S$ is positive Yamabe holds {\it without any caveat} if $\S$ is strictly stably outermost.   To see this, note
that Equation (\ref{der3}) then implies that there exists $\e >0$ such that  
$\left . \frac{\d \th}{\d t} \ \right |_{t=0} \ge \e$.  Tracing through the proof 
using this inequality shows
that  (\ref{psi-ineq})  holds with $Q$ replaced by $Q-\e$.  Then the parenthetical 
expression in Equation~(\ref{scalar})  will include a $+\e$ term, and so $\tilde S$ will be strictly positive.

\medskip
\noindent
2.  Theorem \ref{main} applies, in particular, to the marginally trapped surfaces $S_R$
of a dynamical horizon $\mathcal H$ (see \cite{AK} for definitions).  Indeed, by the
maximum principle for marginally trapped surfaces \cite{AG}, there can be no
outer trapped surfaces in $\mathcal H$ outside of any $S_R$.   Alternatively, it is
easily checked that each $S_R$ is stably outermost in the sense described in the
previous paragraph.

\medskip
\noindent
3.  As discussed in \cite{CG}, the exceptional case in Theorem \ref{main}
can in effect be eliminated in the time symmetric case.  In this case
$V^n$ becomes a manifold of nonnegative scalar curvature, and
$\S^{n-1}$ is minimal.  By the results in 
\cite{CG0,Cai}, if $\S$ is locally outer area minimizing and does not carry
a metric of positive scalar curvature then an outer neighborhood
of $\S$ in $V$ splits isometrically as a product $[0,\e) \times \S$.
In physical terms, this means that there would be marginally outer trapped surfaces
outside of $\S$, which, by a slight strengthening of our definition
of `outer apparent horizon',  could not occur.  (In fact, marginally outer trapped surfaces cannot occur outside the
event horizon.)  Under mild physical assumptions, but with dim $M \le 8$,
one can show that $\S$ is locally outer area minimizing; see \cite[Theorem~3]{CG}
for further discussion.  Finally,  in the  asymptotically flat, but not necessarily 
time symmetric case,
it is possible to perturb the initial data to make the dominant energy
inequality strict, see \cite[p. 240]{SY3}.  Hence, the exceptional case is unstable
in this sense.  

%\section*{Acknowledgments}

\bigskip
\noindent
{\it Acknowledgements.}  This work was supported in part by NSF grants
DMS-0405906 (GJG) and DMS-0104163 (RS).  The work  was initiated
at the Isaac Newton Institute in Cambridge, England during the Fall  2005 Program on Global Problems
in Mathematical Relativity, organized by P. Chru\'sciel, H. Friedrich and
P. Tod.  The authors would like to thank the Newton Institute for its support.

\providecommand{\bysame}{\leavevmode\hbox to3em{\hrulefill}\thinspace}

\end{document}